\documentclass[10pt,twocolumn,twoside]{IEEEtran} 

\IEEEoverridecommandlockouts                              

\usepackage{graphicx,amsmath,amssymb}
\usepackage[noadjust]{cite}
\usepackage{comment,color}
\usepackage{kotex}
\usepackage{algorithm}
\usepackage{algpseudocode}
\usepackage{tikz}
\usepackage{pgfplots}
\usepackage{grffile}
\usepackage{stfloats}
\pgfplotsset{compat=newest}
\usepgfplotslibrary{groupplots}   
\usetikzlibrary{plotmarks}
\usetikzlibrary{arrows.meta}
\usepgfplotslibrary{patchplots}
\usepackage{siunitx}
\usetikzlibrary{shapes,arrows}
\tikzstyle{block} = [draw, fill=white, rectangle, minimum height=3em, minimum width=4em]
\tikzstyle{sum} = [draw, fill=white, circle, node distance=1cm]

\usepackage{enumitem}
\usepackage{amsthm}

\newcommand{\bbR}{\ensuremath{{\mathbb R}}}

\newcommand{\bbN}{\ensuremath{{\mathbb N}}}

\newcommand{\bbP}{\ensuremath{{\mathbb P}}}
\newcommand{\bbE}{\ensuremath{{\mathbb E}}}

\newcommand{\calB}{\mathcal{B}}

\newcommand{\calD}{\mathcal{D}}

\newcommand{\calK}{\mathcal{K}}

\newcommand{\calM}{\mathcal{M}}
\newcommand{\calN}{\mathcal{N}}

\newcommand{\calP}{\mathcal{P}}

\newcommand{\bfzero}{\mathbf{0}}

\newcommand{\nom}{\mathsf{nom}}

\newcommand{\ini}{\mathsf{ini}}

\newcommand{\Adj}{\mathsf{Adj}}

\newcommand{\Lap}{\mathsf{Lap}}

\newcommand{\KL}{\mathsf{KL}}
\newcommand{\Tr}{\mathsf{Tr}}
\newcommand{\logdet}{\log\!\det}

\newtheorem{thm1}{\bf Theorem}
\newtheorem{prop1}{\bf Proposition}
\newtheorem{lem1}{\bf Lemma}
\newtheorem{asm1}{\bf Assumption}
\newtheorem{defn1}{\bf Definition}
\newtheorem{rem1}{\bf Remark}

\newtheorem{cor1}{\bf Corollary}

\newtheorem{prob1}{\bf Problem}

\newenvironment{defn}{\begin{defn1}}{\hfill$\square$\end{defn1}}

\title{\LARGE \bf
A Distributionally Robust Optimal Control Approach for \\ Differentially Private Dynamical Systems
}

\author{Yeongjun Jang, Kaoru Teranishi, and Junsoo Kim
\thanks{*This work was supported by the National Research Foundation of Korea (NRF) grant funded by the Korea government (MSIT) (No. RS-2024-00353032).
}
\thanks{Y.~Jang is with ASRI, Department of Electrical and Computer Engineering, Seoul National University, Seoul, 08826, Korea (email: jangyj0512@snu.ac.kr).
}
\thanks{K.~Teranishi is with the Department of Information and Physical Sciences, Graduate School of Information Science and Technology,
The University of Osaka, Osaka, 565-0871, Japan (email: k-teranishi@ist.osaka-u.ac.jp).
}
\thanks{J.~Kim is with the Department of Electrical and Information Engineering,
Seoul National University of Science and Technology, Seoul, 01811, Korea (email: junsookim@seoultech.ac.kr).
}
}

\begin{document}

\maketitle
\thispagestyle{plain} 
\pagestyle{plain} 



\begin{abstract}
In this paper, we develop a distributionally robust optimal control approach for differentially private dynamical systems, enabling a plant to securely outsource control computation to an untrusted remote server. 
We consider a plant that ensures differential privacy of its state trajectory by injecting calibrated noise into its output measurements.
Unlike prior works, we assume that the server only has access to an ambiguity set consisting of admissible noise distributions, rather than the exact distribution. 
To account for this uncertainty, the server formulates a distributionally robust optimal control problem to minimize the worst-case expected cost over all admissible noise distributions.
However, the formulated problem is computationally intractable due to the nonconvexity of the ambiguity set.
To overcome this, we relax it into a convex Kullback--Leibler divergence ball, so that the reformulated problem admits a tractable closed-form solution.

\end{abstract}

\section{Introduction}

The advancement of cloud computing has enabled resource-limited devices to outsource computationally intensive tasks to remote servers, thereby improving scalability and efficiency \cite{LimhBabu09,HegaHefe14,Liug17}. 
However, such delegation requires transmitting data that may contain sensitive information (e.g., current state or model parameters), leading to privacy concerns. 
In particular, the data sent over communication channels are vulnerable to eavesdropping and the server may be semi-honest, meaning that it correctly executes the assigned protocol while attempting to infer sensitive information. 
Therefore, the problem of preserving data utility while providing formal privacy guarantees has attracted significant interest.

Recently, differential privacy (DP) has emerged as a powerful tool for preserving both data utility and privacy \cite{DworMcsh06,DworRoth14}.
Rather than releasing raw data, DP adds calibrated noise, so that adversaries cannot accurately infer the input data from the noisy (privatized) output data. 
It has been widely adopted across various applications due to several appealing features.
In particular, its immunity to post-processing and resilience to side information ensures that the privacy guarantees are preserved under arbitrary manipulation of the released output and when an adversary possesses auxiliary knowledge \cite{DworRoth14}.

In the control literature, DP has typically been utilized to privatize a plant's state trajectory by injecting artificial noise into the input and/or output, followed by the synthesis of an optimal filter \cite{LenyPapp13,CortDull16,DeguLeny22b} or an optimal controller \cite{HaleJone18,YazdJone22,DeguLeny22a} to mitigate the effect of noise.
In particular, the aforementioned works restrict their attention to injecting Gaussian noise, allowing them to directly apply the standard Kalman filter or linear quadratic Gaussian (LQG) control.
This setting, however, entails two key limitations. 
First, injecting Gaussian noise can only guarantee a weaker notion of DP (see Section~\ref{sec:DP}), which may be inadequate in privacy-sensitive applications. 
While a stronger notion can be achieved by employing suitable non-Gaussian noises (e.g., Laplace noise), doing so makes the Kalman filter or LQG control fundamentally inapplicable.
Second, the Kalman filter or LQG control require exact knowledge of the noise statistics.
This can be particularly problematic in cloud based control settings, in which the plant may be unwilling to disclose these parameters to the server due to privacy concerns.

To overcome these limitations, we develop a distributionally robust optimal control approach for differentially private dynamical systems. 
We consider a cloud based control setting in which the plant ensures DP of its state trajectory by adding either Gaussian or Laplace noise to its output, and the server only has access to an admissible range of the noise parameters. 
To guarantee robust performance, the server formulates a distributionally robust optimal control problem to synthesize an output feedback controller that minimizes the worst-case expected cost over all admissible noise distributions. 

The resulting problem, however, is computationally intractable as the underlying ambiguity set formed by the union of Gaussian and Laplace distributions is nonconvex.
To address this, we construct a convex Kullback-Leibler divergence ball that contains all admissible noise distributions and relax the original ambiguity set.
This relaxation allows for a reformulation into a risk-sensitive control problem that admits a computationally tractable closed-form solution at the cost of suboptimality.
\textit{To the best of our knowledge, this is the first result to synthesize an optimal controller for differentially private dynamical systems while accounting for both non-Gaussian noise and distributional ambiguity.}

\textit{Notations:}
Let $\bbR$ and $\bbN$ denote the sets of real numbers and positive integers.
For a sequence $v_1,\ldots,v_n$ of scalars or vectors, we define $v_{1:n} := [v_1^\top, \ldots, v_n^\top]^\top$.
The identity and the zero matrices are denoted by $I$ and $\bfzero$, respectively, with their dimensions indicated as subscripts when necessary.
For a probability distribution $P$ (or a random variable $X\sim P$), we denote its probability density function by $\pi_P$ (or $\pi_X)$.
We write $X\sim \calN(\mu,\Sigma)$ to denote that a random variable $X$ follows a multivariate Gaussian distribution with mean $\mu$ and covariance matrix $\Sigma$.
Similarly, we use $X \sim \Lap(b,h)$ to denote an $\bbR^h$-valued random vector whose elements each independently follows a zero-mean Laplace distribution with the scale parameter $b>0$.

\section{Preliminaries and Problem Formulation}

\subsection{Differential privacy of dynamical systems}\label{sec:DP}

We introduce the basic notions of differential privacy (DP), specifically adapted to dynamical systems.
The core idea is to inject calibrated measurement noise such that the output trajectories generated from adjacent state trajectories are nearly indistinguishable, thereby preventing accurate inference of the underlying state based on the observed output.

To formalize this, consider a discrete-time plant written by 
\begin{align}\label{eq:plant}
    x(k+1) &= A x(k) + Bu(k) + w(k), \\
    y(k) &= Cx(k), \nonumber
\end{align}
where $x(k) \in \bbR^n$ is the state, $y(k) \in \bbR^p$ is the output, $u(k) \in \bbR^m$ is the input, and $w(k)\in\bbR^n$ is the process noise. 

To privatize its state trajectory $x_{0:N}$ over a fixed horizon $N\in\bbN$, the plant conceals the raw output $y(k)$.
Instead, it publishes a privatized output
\begin{align*}
    \tilde{y}(k) = y(k) + v(k)
\end{align*}
by injecting an artificial measurement noise $v(k)\in\bbR^p$ at each time step $k$.
Accordingly, the plant can be modeled as a randomized mechanism $\calM$, defined as
\begin{align}\label{eq:single_mech}
    \calM(x_{0:N}):=\tilde{y}_{0:N}\in\bbR^{L},
\end{align}
where the randomness arises from $v_{0:N}$ and $L=p(N+1)$.

Let $\calD:=\bbR^{n(N+1)}$ denote the set of all state trajectories of length $N+1$.
The set of all adjacent state trajectory pairs is defined as 
\begin{align}\label{eq:adjDef}
        \Adj \!:=\! \left\{ (x_{0:N}, x'_{0:N}) \!\in\! \calD \times \calD \mid \|x_{0:N} - x'_{0:N}\|_{1} \!\le\! \gamma \right\}
    \end{align}
for some tunable parameter $\gamma > 0$.
This definition is natural in the sense that two state trajectories are considered adjacent if their $\ell_1$-distance is less than or equal to $\gamma$.

Based on these, $(\epsilon,\delta)$-DP and $\epsilon$-DP are defined as follows.

\begin{defn1}\upshape
    Let the plant \eqref{eq:plant} be modeled as a randomized mechanism $\calM$ as defined in \eqref{eq:single_mech}.
    For given $\epsilon\ge 0$ and $\delta\in[0,1)$, the plant satisfies $(\epsilon,\delta)$-DP if
    \begin{align*}
        \bbP\left[\calM(x_{0:N})\in S \right] \le \exp(\epsilon)\cdot \bbP\left[\calM(x_{0:N}')\in S \right] + \delta
    \end{align*}
    for all $S\subseteq\bbR^{L}$ and $(x_{0:N},x_{0:N}')\in\Adj$. 
    If $\delta=0$, the plant is said to satisfy $\epsilon$-DP.
\end{defn1}

The parameter $\epsilon$, often referred to as the \textit{privacy budget}, governs the maximum allowable change in the output probability distribution for adjacent inputs, with smaller $\epsilon$ corresponding to stronger privacy.
The parameter $\delta$ relaxes this by allowing a failure probability of at most $\delta$. 
Consequently, $\epsilon$-DP is a stronger notion and implies $(\epsilon,\delta)$-DP for any $\delta\ge 0$.

An appealing feature of DP is its immunity to post-processing. 
That is, applying any transformation on the mechanism's output, such as for feedback control or state estimation, does not degrade the established privacy level.
Additionally, DP admits an additive composition rule that facilitates the characterization of cumulative privacy loss incurred over time.
For a comprehensive treatment of these properties and other aspects of DP, we refer the reader to \cite{DworRoth14}.

In what follows, we introduce two representative and widely used mechanisms for ensuring DP; the Gaussian and Laplace mechanisms.
The Gaussian mechanism ensures $(\epsilon,\delta)$-DP by drawing $v(k)$ from a multivariate Gaussian distribution.
While it is limited to $(\epsilon, \delta)$-DP, the Laplace mechanism can ensure the stronger $\epsilon$-DP by using the Laplace distribution.

\begin{lem1}\label{lem:mechGauss}\upshape
    Let the plant \eqref{eq:plant} be modeled as a randomized mechanism $\calM$ as defined in \eqref{eq:single_mech}.
    \begin{enumerate}
        \item For given $\epsilon\in(0,1)$ and $\delta\in(0,1)$, if $v_{0:N}\sim\calN(\bfzero,\sigma^2 I)$ with
        \begin{align*}
        \sigma^2 &\ge \frac{2\ln(1.25/\delta)}{\epsilon^2} \left\| C\right\|_2^2\gamma^2,
        \end{align*}
        then the plant satisfies $(\epsilon,\delta)$-DP. 
        \item For given $\epsilon> 0$, if $v_{0:N}\sim\Lap(b,L)$ with
        \begin{align*}
        b &\ge \frac{\left\|C \right\|_1 \gamma}{\epsilon},
    \end{align*}
        then the plant satisfies $\epsilon$-DP. 
    \end{enumerate}
\end{lem1}
\begin{proof}
    See Appendix~\ref{apdx:dp}.
\end{proof}

The derived lower bounds for the variance $\sigma^2$ and the scale parameter $b$ are inversely proportional to $\epsilon^2$ and $\epsilon$, respectively. 
This implies that achieving stronger privacy necessitates injecting noise with larger variance or scale.
Also, observe that the lower bounds scale with $\gamma$, indicating that protecting privacy across a larger $\Adj$ requires noise with larger variance or scale.
Conversely, these bounds decrease as the gain $\|C\|_1$ decreases, which suggests that systems with lower sensor sensitivity inherently render the output trajectories harder to distinguish.

\subsection{Problem formulation}\label{sec:problem}

We consider a cloud based control architecture in which the plant \eqref{eq:plant} transmits the privatized output $\tilde{y}(k)$ to a semi-honest remote server.
Based on the received output history, the server computes and returns the control input $u(k)$, while simultaneously attempting to infer the underlying state trajectory $x_{0:N}$.
To protect its state trajectory, the plant designs the noise sequence $v_{0:N}$ using either the Gaussian or Laplace mechanism to satisfy a desired $(\epsilon,\delta)$-DP guarantee.

Crucially, unlike prior works \cite{LenyPapp13,CortDull16,DeguLeny22b,HaleJone18,YazdJone22,DeguLeny22a}, we assume that neither the specific mechanism nor the noise parameters chosen by the plant are known to the server, as the plant may be unwilling to share such information due to privacy concerns.
Instead, we assume that the server has access to an ambiguity set $\Xi$ consisting of admissible noise distributions, defined as
\begin{align*}
    \Xi :=  \left\{\calN(\bfzero,\sigma^2 I) \mid \sigma^2 \in [\underline{\sigma}^2,\overline{\sigma}^2] \right \}
    \cup \left\{\Lap(b,L)\mid  b \in [\underline{b},\overline{b}] \right\}.
\end{align*}
That is, the server only knows that $v_{0:N}\sim P$ for some unknown $P\in\Xi$. 
Here, the lower bounds $\underline{\sigma}^2>0$ and $\underline{b}>0$ are chosen to satisfy the conditions in Lemma~\ref{lem:mechGauss}, and $\overline{\sigma}^2>0$ and $\overline{b}>0$ are empirical upper bounds introduced to prevent the noise parameters from being chosen excessively large.

Given $\Xi$, the server aims to synthesize an optimal controller from the set of admissible controllers $\Lambda$, defined as 
\begin{align*}
    \Lambda \subseteq \left\{\calK=\{\calK_k\}_{k=0}^{N-1} \mid \calK_k:\bbR^{p(k+1)} \to \bbR^m \right\}.
\end{align*}
That is, any $\calK\in\Lambda$ is a causal output feedback controller that generates the control input as $u(k) = \mathcal{K}_k(\tilde{y}_{0:k})$.
Since the exact distribution of $v_{0:N}$ remains unknown, standard optimal control methods, such as LQG, are fundamentally inapplicable.

To overcome this challenge, the server formulates a distributionally robust optimal control problem.
Specifically, the goal is to find a controller $\calK\in\Lambda$ that minimizes the worst-case expected cost functional over $\Xi$, thus guaranteeing robust performance against any admissible noise distribution.
The problem of interest is formally stated as follows.

\begin{prob1}\label{prob}\upshape
    For the plant \eqref{eq:plant}, assume that the initial state satisfies $x(0)\sim\calN(x_\ini,\Sigma_\ini)$ for some $x_\ini\in\bbR^n$ and $\Sigma_\ini\succ 0$, and that the process noise is white Gaussian with $w(k)\sim\calN(\bfzero,\Sigma_w)$, where $\Sigma_w\succ 0$. 
    
    Given the parameters $\{A,B,C, x_\ini, \Sigma_\ini, \Sigma_w, N\}$ and the ambiguity set $\Xi$, design a controller $\calK\in\Lambda$ by solving the following distributionally robust optimal control problem:
    \begin{align}\label{eq:minimax_prob1}
        \inf_{\mathcal{K} \in \Lambda} \sup_{P\in \Xi} \bbE_{v_{0:N}\sim P} \left[ J(\mathcal{K}) \right],
    \end{align}
    where the expectation is taken jointly\footnote{For notational brevity, the dependency on $x_0$ and $w_{0:N-1}$ is omitted.} over $x(0)$, $w_{0:N-1}$, and $v_{0:N}$, and the finite-horizon cost $J(\calK)$ is defined as
    \begin{multline*}
        J(\calK) :=\frac{1}{2}x(N)^\top Q_N x(N) \\
        + \frac{1}{2} \sum_{k=0}^{N-1} \left( x(k)^\top Q x(k) + u(k)^\top R u(k)\right)
    \end{multline*}
    with $Q_N\succeq 0$, $Q\succeq 0$, and $R\succ 0$. 
\end{prob1}

Before proceeding, we impose the following assumption on $\Lambda$, which has also been made in \cite[Assumption~3.2]{PeteJame02}.
\begin{asm1}\label{asm:ctrb}\upshape
    For any admissible controller $\calK\in\Lambda$, 
    \begin{align*}
        \sup_{P\in\calP(\bbR^L)} \bbE_{v_{0:N}\sim P} \left[J(\calK)\right] = \infty,
    \end{align*}
    where $\calP(\bbR^L)$ is the set of all probability distributions on $\bbR^{L}$.
\end{asm1}

Assumption~\ref{asm:ctrb} implies that for any admissible controller $\calK\in\Lambda$, the associated expected cost can be made arbitrarily large by suitably choosing the distribution of $v_{0:N}$ to sufficiently corrupt $\tilde{y}_{0:N}$.
Hence, $\Lambda$ excludes degenerate controllers that ignore the output history, for example, constant controllers.

\section{Main Results}\label{sec:Main}

\subsection{Tractable reformulation of the optimization problem}

Directly solving the minimax optimization problem \eqref{eq:minimax_prob1} is computationally intractable due to the nonconvex nature of the ambiguity set $\Xi$, which is formed as a union of Gaussian and Laplace distributions. 
Indeed, a convex combination of a Gaussian and a Laplace distribution is generally neither Gaussian nor Laplace.
To address this, we relax the ambiguity set $\Xi$ into a convex Kullback-Leibler (KL) divergence ball, which results in a tractable reformulation of \eqref{eq:minimax_prob1}.

Formally, the KL divergence is defined as follows.
\begin{defn}\label{def:KL}\upshape
    Let $P$ and $Q$ be two probability distributions on $\bbR^L$.
    The KL divergence of $P$ from $Q$ is defined as 
    \begin{align*}
        D_\KL(P\| Q) := \int_{\bbR^L} \pi_P(x) \log \left( \frac{\pi_P(x)}{\pi_Q(x)}\right) dx.
    \end{align*}
    If there exists $x\in\bbR^L$ such that $\pi_P(x)>0$ but $\pi_Q(x)=0$, we define $D_\KL(P\| Q)=\infty$. 
\end{defn}

Let us fix $P_\nom := \calN(\bfzero,\underline{\sigma}^2 I)\in \Xi$
as our \textit{nominal} probability distribution. 
We first derive an explicit closed-form expression for the KL divergence between a Laplace distribution and $P_\nom$.

\begin{lem1}\label{lem:lap}\upshape
    The KL divergence of $P\sim\Lap(b,L)$ from $P_\nom$ is given by 
\begin{align}\label{eq:lemGaussToShow}
         D_\KL (P\| P_\nom) 
        = \frac{L}{2}\left( \log \frac{\underline{\sigma}^2}{2b^2} + \frac{2b^2}{\underline{\sigma}^2}-2+\log\pi\right).
    \end{align}
\end{lem1}

\begin{proof}
    The KL divergence can be alternatively expressed as the difference between the expected log-likelihoods, as
    \begin{align*}
        D_\KL(P\| P_\nom) = \bbE_{x\sim P}\left[ \log \pi_P(x)\right] - \bbE_{x\sim P}\left[ \log\pi_{P_\nom}(x)\right].
    \end{align*}
    The first term is the negative entropy of the Laplace distribution $P$, which is given by \cite[Chapter~2.1]{KotzKozu12}
    \begin{align}\label{eq:lemGauss1}
        \bbE_{x\sim P}\left[ \log\pi_P(x)\right] = L\left(-1+\log \frac{1}{2b} \right) .
    \end{align}
    
    For the second term, expanding the log-likelihood of the nominal Gaussian density $\pi_{P_\nom}$ yields
    \begin{align}\label{eq:lemGauss2}
        -\bbE_{x\sim P}\left[ \log\pi_{P_\nom}(x)\right] &= \frac{L}{2}\log\pi + \frac{L}{2} \log(2\underline{\sigma}^2) \\
        &+\frac{1}{2\underline{\sigma}^2}\bbE_{x\sim P}\left[x^\top x\right]. \nonumber
    \end{align}
    To evaluate the expectation of the quadratic term, we utilize the fact that $\bbE_{x\sim P}[x] =\bfzero$ and $\bbE_{x\sim P}[xx^\top] = 2b^2I_L $ \cite[Chapter~2.1]{KotzKozu12}, which leads to
    \begin{align*}
        \bbE_{x\sim P}\left[x^\top x\right] = \Tr\left(\bbE_{x\sim P}\left[xx^\top \right]  \right)= \Tr\left( 2b^2 I_L\right) = 2b^2L,
    \end{align*}
    where $\Tr(\cdot)$ is the trace operator. 
    Combining this with \eqref{eq:lemGauss2} and \eqref{eq:lemGauss1} results in \eqref{eq:lemGaussToShow}, and this concludes the proof.
\end{proof}

Building on Lemma~\ref{lem:lap}, the following theorem establishes a KL divergence ball centered at $P_\nom$ that contains all admissible noise distributions in $\Xi$.

\begin{thm1}\label{thm:ball}\upshape

For any $P\in\Xi$, the KL divergence of $P$ from $P_\nom$ is bounded as
\begin{align}\label{eq:deviationBound}
    D_{\KL}(P \| P_\nom  )\le \frac{L}{2}\max\left\{ \eta_1, \eta_2 \right\}
    =:\eta>0,
\end{align}
where
\begin{align*}
    \eta_1 &:= g(\overline{\sigma}^2) - 1\in\bbR, \\
    \eta_2 &:= \max\left\{g(2\underline{b}^2),g(2\overline{b}^2) \right\}-2+\log\pi \in \bbR
\end{align*}
with $g(x) = \log(\underline{\sigma}^2/x) + x/\underline{\sigma}^2$.
\end{thm1}

\begin{proof}
First, suppose $P=\calN(\bfzero,\sigma^2  I_L)$ with $\sigma^2 \in [\underline{\sigma}^2,\overline{\sigma}^2]$. 
Using the standard closed-form expression for the KL divergence between two Gaussian distributions \cite{GilmAlaj13}, we have
\begin{align}\label{eq:gaussKLMax}
&D_{\KL}(P\| P_{\nom}) \\
&= \frac{1}{2} \bigg(\log\left(\frac{\det(\underline{\sigma}^2 I_L)}{\det(\sigma^2 I_L)} \right) +\Tr\left( \frac{\sigma^2}{\underline{\sigma}^2} I_L \right) - L \bigg) \nonumber \\
&= \frac{1}{2} \left( L\log\left(\frac{\underline{\sigma}^2}{\sigma^2} \right) + L \frac{\sigma^2}{\underline{\sigma}^2} - L \right)  \nonumber \\
&= \frac{L}{2}\left( g(\sigma^2) - 1  \right) \nonumber\le \frac{L}{2} \eta_1,
\end{align}
where the last inequality follows from the fact that $g(x)$ is increasing for $x\ge \underline{\sigma}^2$.

Next, suppose that $P=\Lap(b,L)$ with $b\in[\underline{b},\overline{b}]$. 
By applying Lemma~\ref{lem:lap}, it is obtained that
\begin{align}\label{eq:lapKL}
    D_\KL(P \| P_\nom)=\frac{L}{2}\left(g(2b^2) - 2 + \log\pi\right). 
\end{align}
Observe that $g(x)$ is strictly convex on $(0,\infty)$ and attains a unique global minimum at $x=\underline{\sigma}^2$. 
Therefore, the right-hand-side of \eqref{eq:lapKL} is maximized at one of the boundary points of $[\underline{b},\overline{b}]$, and thus, $D_\KL(P \| P_\nom) \le L\eta_2/2$.
Combining this with \eqref{eq:gaussKLMax} leads to \eqref{eq:deviationBound}.
Moreover, since $g(\underline{\sigma}^2)=1$, it holds that $\eta_2\ge \log \pi - 1>0$, implying $\eta>0$.
This concludes the proof.
\end{proof}

Based on Theorem~\ref{thm:ball}, we construct an ambiguity set $\calB_\eta$ as the KL divergence ball of radius $\eta$ centered at $P_\nom$:
\begin{align*}
    \calB_\eta := \left\{ P\in\calP(\bbR^L) \mid D_{\KL}( P \| P_\nom) \le \eta\right\}.
\end{align*}
Since $\Xi\subset \calB_\eta$ by construction, we can relax \eqref{eq:minimax_prob1} by replacing the ambiguity set $\Xi$ with $\calB_\eta$, leading to the reformulated problem
\begin{align}\label{eq:minimax_prob2}
    \inf_{\mathcal{K} \in \Lambda} \sup_{P \in \calB_\eta} \bbE_{v_{0:N}\sim P} \left[ J(\mathcal{K}) \right].
\end{align}

While \eqref{eq:minimax_prob2} provides a suboptimal solution to \eqref{eq:minimax_prob1}, we emphasize that its inner maximization is now a convex optimization problem in $P$ for any fixed $\calK\in\Lambda$.
This is because $\calB_\eta$ is convex \cite[Chapter~3]{BoydVand04} and the expected cost is affine in $P$.

In the following subsection, we show that the reformulated problem \eqref{eq:minimax_prob2} is closely related to the risk-sensitive optimal control problem, for which well-established solutions exist. 

\begin{rem1}\label{rem:center}\upshape
    Our specific choice of $P_\nom$ is for analytical convenience, and theoretically, it may be chosen as any alternative distribution.
    Selecting a different $P_\nom$ might yield a tighter radius $\eta$, thereby reducing the conservatism of the synthesized controller.
    However, optimizing the choice of $P_\nom$ is beyond the scope of this work.
\end{rem1}

\subsection{Control design}

The following theorem establishes an equivalent representation of \eqref{eq:minimax_prob2} whose inner optimization problem reduces to the standard risk-sensitive optimal control problem \cite{Whit81,CollJame96}.

\begin{thm1}\label{thm:problEquiv}\upshape
Under Assumption~\ref{asm:ctrb}, the minimax optimal control problem \eqref{eq:minimax_prob2} is equivalent to
\begin{align}\label{eq:equiv}
    \inf_{\tau>0} \tau\left(\eta  + W_\tau\right),
\end{align}
where $W_\tau$ denotes the optimal value of a risk-sensitive optimal control problem, given by
\begin{align} \label{eq:risk_sensitive}
W_\tau = \inf_{\calK\in\Lambda}  \log \bbE_{v_{0:N}\sim P_\nom}\left[ \exp \left(\frac{J(\calK)}{\tau} \right) \right] .
\end{align}
\end{thm1}

\begin{proof}
    For any fixed $\calK\in\Lambda$, consider the inner maximization problem of \eqref{eq:minimax_prob2}, which is a convex optimization problem.
    Since $P_\nom \in \calB_\eta$ is strictly feasible, i.e., $D_\KL(P_\nom\| P_\nom)=0 < \eta$, strong duality holds by Slater's condition \cite[Chapter~5]{BoydVand04}, and thus,
    \begin{align}\label{eq:strong}
        &\sup_{P \in \calB_\eta} \bbE_{v_{0:N}\sim P} \left[ J(\mathcal{K}) \right] \\
        &=\inf_{\tau \ge 0} \sup_{P\in \calP(\bbR^L)} \bbE_{v_{0:N}\sim P} \left[J(\calK) - \tau\left(D_{\KL}(P\| P_\nom) - \eta \right) \right] \nonumber \\
        &= \inf_{\tau > 0} \sup_{P\in \calP(\bbR^L)} \bbE_{v_{0:N}\sim P} \left[J(\calK) - \tau\left(D_{\KL}(P\| P_\nom) - \eta \right) \right], \nonumber
    \end{align}
    where the second equality follows from Assumption~\ref{asm:ctrb}.
    For any $\tau>0$, the inner maximization problem of the right-hand-side of \eqref{eq:strong} can be rewritten as 
    \begin{align*}
        &\sup_{P\in \calP(\bbR^L)} \bbE_{v_{0:N}\sim P} \left[J(\calK) - \tau\left(D_{\KL}(P\| P_\nom) - \eta \right) \right] \\
        &=\tau\eta + \tau\sup_{P\in \calP(\bbR^L)} \left( \bbE_{v_{0:N}\sim P}\left[\frac{J(\calK)}{\tau}\right] - D_{\KL}(P\| P_\nom) \right) \\
        &=\tau\eta + \tau \log \bbE_{v_{0:N}\sim P_\nom} \left[ \exp\left( \frac{J(\calK)}{\tau} \right) \right],
    \end{align*}
    where the last equality follows from the Donsker-Varadhan variational formula \cite{DupuElli11}.
    Substituting this into \eqref{eq:strong} concludes the proof.
\end{proof}

In the literature, \eqref{eq:risk_sensitive} is widely recognized as the risk-sensitive optimal control problem, where $\theta:=1/\tau$ represents the risk-sensitivity parameter. 
Since $\tau>0$, we have $\theta>0$, which corresponds to a risk-averse regime. 
That is, the exponential transformation in  \eqref{eq:risk_sensitive} assigns heavier weight to tail outcomes with high cost, enforcing robustness against worst-case noise realizations.

For a fixed $\tau>0$, the optimal control policy for \eqref{eq:risk_sensitive} admits an LQG-like structure \cite{Whit81,CollJame96}.
Specifically, it consists of a state estimator and a feedback policy that are coupled through the parameter $\tau$.
Since existing results typically account for distributional ambiguity across the initial state, process noise, and measurement noise, we adapt them to our setting in which the ambiguity is confined to the measurement noise.

The state estimator is characterized by the forward Riccati equation written by 
\begin{align*}
    \Sigma_{k+1} = \Sigma_w + AP_k ^{-1}A^\top,~~~P_k := \Sigma_k^{-1} +\frac{C^\top C}{\underline{\sigma}^{2}}- \frac{Q}{\tau},
\end{align*}
initialized at $\Sigma_0=\Sigma_\ini$, provided that $P_k\succ 0$ and $\Sigma_k\succ 0$ for $k=0,\ldots,N-1$.
Intuitively, this condition could fail when $\tau$ is sufficiently small, i.e., when $\theta$ is sufficiently large, implying a maximum threshold on the achievable risk-sensitivity.
The dynamics of the state estimate $\hat{x}(k)\in\bbR^n$ is then given by 
\begin{align*}
    \hat{x}(k+1) &= A\hat{x}(k) + Bu(k) + K_k\left(\tilde{y}(k)-C\hat{x}(k) \right) \\
    &+ \frac{ AP_k^{-1}Q }{\tau}\hat{x}(k),~~~\hat{x}(0)=x_\ini,
\end{align*}
where the gain is defined as $K_k=AP_k^{-1}C^\top/\underline{\sigma}^2$.

The feedback policy is determined by the backward Riccati equation written by
\begin{align*}
    \Pi_k = Q + A^\top L_{k+1}^{-1}A,~~~
    L_{k+1} :=  \Pi_{k+1}^{-1} +BR^{-1}B^\top - \frac{\Sigma_w}{\tau},
\end{align*}
initialized at $\Pi_{N}=Q_{N}$. 
To ensure the existence of a stabilizing feedback gain, the solution is required to satisfy $\Pi_{k+1}^{-1} - \Sigma_w/\tau \succ 0$ and $\Pi_k^{-1}-\Sigma_k/\tau \succ 0$ for $k=0,\ldots, N-1$, similar to the conditions implied on the state estimator.

The following proposition provides closed-form expressions for the optimal value and the associated optimal policy of \eqref{eq:risk_sensitive}.
The proof can be obtained by adapting the results of \cite{Whit81,CollJame96}, and is therefore omitted here due to space limitations.

\begin{prop1}\upshape\label{prop:controller}
    For a fixed $\tau>0$, let $\calK^\tau\in\Lambda$ denote the optimal policy for the risk-sensitive optimal control problem \eqref{eq:risk_sensitive}.
    The optimal value $W_\tau$ is given by \eqref{eq:optVal} and the optimal control input $u(k)=\calK^\tau_k(\tilde{y}_{0:k})$ is given by
    \begin{align*}
        u(k) = -R^{-1}B^\top L_{k+1}^{-1}A\left(I-\frac{\Sigma_k\Pi_k}{\tau} \right)^{-1} \hat{x}(k)
    \end{align*}
    for $k=0,\ldots,N-1$.
\end{prop1}

This proposition enables us to reduce \eqref{eq:equiv} to an outer optimization over $\tau>0$.
Since a closed-form expression for the optimal $\tau^*$ is generally unavailable, it is standard in practice to determine it by evaluating the objective in \eqref{eq:equiv} over feasible $\tau$ by utilizing the closed-form expression for $W_\tau$ \cite{Pete06,PeteJame02}, as illustrated in Fig.~\ref{fig:tau}.
\begin{figure}[t]
    \centering
\begin{tikzpicture}

\definecolor{darkgray176}{RGB}{176,176,176}
\definecolor{lightgray204}{RGB}{204,204,204}
\definecolor{steelblue31119180}{RGB}{31,119,180}
\definecolor{steelblue}{RGB}{70,130,180}

\begin{axis}[
width=0.47\textwidth,
height=0.15\textwidth,
tick align=outside,
tick pos=left,
x grid style={darkgray176},
xlabel={$\tau$},
xmajorgrids,
xmin=15, xmax=100,
xtick style={color=black},
xlabel style={yshift=3.5mm},
y grid style={darkgray176},
ylabel={$\tau(\eta+W_\tau)$},
ymajorgrids,
ymin=100, ymax=250,
ytick={100,200},
ytick style={color=black},
extra x ticks={28.1392},
extra x tick labels={$\tau^*$},
extra x tick style={tick label style={text=red}},
label style={font=\footnotesize},      
tick label style={font=\footnotesize}, 
title style={font=\footnotesize}
]
\addplot [line width=1.2pt, steelblue]
table {%
20.0275135294006 186.71947312454
20.2014120539119 153.527829032743
20.3768205360513 144.216018237103
20.5537520867527 138.927183778273
20.7322199307918 135.28719645596
20.9122374077748 132.534983283195
21.0938179731357 130.335544377775
21.2769751991414 128.513420616861
21.4617227759068 126.965796514922
21.6480745124177 125.627462453987
21.8360443375631 124.454544042489
22.0256463011763 123.41613619454
22.216894575085 122.489655422898
22.4098034541706 121.65809535374
22.6043873574368 120.908323974251
22.8006608290871 120.229983069635
22.9986385396121 119.614751931293
23.198335286886 119.055840038914
23.3997659972727 118.547628486192
23.6029457267414 118.085410829357
23.8078896619919 117.665202074288
24.0146131215901 117.28359540517
24.2231315571124 116.937653028112
24.4334605543012 116.624821826252
24.6456158342294 116.342867349415
24.8596132544759 116.08982154897
25.0754688103105 115.863940953605
25.2931986358896 115.663672872115
25.5128190054624 115.487627835871
25.7343463345869 115.334556941411
25.9577971813571 115.203333077584
26.1831882476406 115.092935259292
26.4105363803272 115.00243546597
26.6398585725877 114.930987515042
26.8711719651444 114.877817600434
27.1044938475522 114.842216202668
27.3398416594907 114.823531135872
27.5772329920681 114.821161542779
27.8166855891357 114.834552684619
28.0582173486147 114.863191401041
28.301846323833 114.906602137661
28.5475907248758 114.964343456733
28.7954689199457 115.036004960912
29.0454994367363 115.121204571695
29.2977009638168 115.219586113661
29.5520923520289 115.330817163385
29.8086926158955 115.454587128245
30.0675209350428 115.590605525681
30.3285966556328 115.738600437741
30.5919392918102 115.898317119492
30.8575685271605 116.069516742836
31.1255042161815 116.251975259887
31.3957663857672 116.445482372179
31.6683752367047 116.649840593847
31.9433511451841 116.864864398445
32.2207146643217 117.090379440378
32.5004865256963 117.326221843099
32.782687640898 117.572237547132
33.0673391030926 117.828281711895
33.3544621885969 118.094218165906
33.6440783584698 118.369918900712
33.9362092601161 118.655263604299
34.2308767289048 118.9501392303
34.5281027898006 119.254439599667
34.827909659011 119.568065031876
35.1303197456462 119.890922003025
35.4353556533942 120.222922828457
35.7430401822105 120.56398536781
36.0533963300222 120.914032750573
36.366447294447 121.272993120452
36.682216474527 121.640799397001
37.0007274724778 122.017389053104
37.3220040954525 122.402703907082
37.6460703573214 122.79668992824
37.9729504804667 123.19929705486
38.3026688975931 123.610479023651
38.635250253554 124.03019320984
38.9707194071937 124.458400477089
39.3091014332052 124.895065036538
39.6504216240049 125.340154314318
39.9947054916223 125.793638826937
40.3419787696078 126.255492063986
40.6922674149556 126.725690377668
41.045597610044 127.204212878684
41.4019957645922 127.691041338048
41.7614885176349 128.186160094451
42.1241027395126 128.6895559668
42.4898655338808 129.201218171607
42.8588042397354 129.721138244915
43.2309464334563 130.249309968482
43.6063199308686 130.785729299961
43.9849527893216 131.33039430682
44.3668733097861 131.883305103788
44.7521100389698 132.444463793618
45.1406917714508 133.013874410959
45.5326475518301 133.591542869172
45.9280066769022 134.177476909898
46.3267986978453 134.771686055262
46.7290534224299 135.374181562509
47.1348009172469 135.984976380993
47.5440715099546 136.604085111345
47.9568957915461 137.231523966734
48.3733046186352 137.867310736092
48.7933291157633 138.511464749192
49.2170006777256 139.164006843515
49.6443509719177 139.824959332771
50.0754119407025 140.494345977015
50.510215803798 141.172191954275
50.9487950606853 141.858523833609
51.3911824930378 142.553369549525
51.8374111671719 143.256758377695
52.2875144365179 143.968720911905
52.7415259441133 144.689289042169
53.1994796251176 145.418495933977
53.6614097093486 146.156376008591
54.1273507238409 146.902964924367
54.5973374954269 147.658299559047
55.0714051533396 148.422417992975
55.5495891318384 149.19535949319
56.0319251728577 149.977164498381
56.5184493286786 150.767874604633
57.0091979646234 151.567532551958
57.5042077617739 152.376182211561
58.0035157197129 153.193868573818
58.5071591592899 154.02063773693
59.0151757254107 154.856536896232
59.5276033898511 155.701614334126
60.0444804540952 156.555919410614
60.565845552198 157.419502554408
61.0917376536734 158.29241525459
61.6221960664067 159.174710052808
62.1572604395931 160.066440535984
62.6969707667007 160.967661329512
63.2413673884604 161.878428090931
63.7904909958809 162.798797504056
64.3443826332899 163.728827273546
64.9030837014025 164.668576119904
65.4666359604153 165.618103774875
66.0350815331279 166.577470977251
66.6084629080915 167.546739469051
67.1868229427846 168.525971992069
67.770204866816 169.515232284785
68.3586522851569 170.514585079617
68.9522091813994 171.524096100505
69.5509199210443 172.543832060824
70.1548292548172 173.573860661603
70.7639823220136 174.614250590056
71.3784246538723 175.665071518403
71.9982021769793 176.726394102973
72.6233612167003 177.798289983593
73.253948500643 178.880831783245
73.8900111621503 179.974093107975
74.5315967438229 181.078148547072
75.1787532010731 182.193073673483
75.831528905709 183.318945044477
76.4899726495505 184.455840202539
77.1541336480756 185.603837676498
77.8240615440996 186.763016982874
78.4998064114853 187.933458627448
79.181418758886 189.11524410704
79.8689495335208 190.308455911498
80.5624501249822 191.513177525892
81.261972369078 192.729493432902
81.967568551705 193.957489115402
82.6792914127576 195.197251059242
83.3971941500698 196.448866756203
84.1213304233911 197.712424707145
84.8517543583981 198.988014425334
85.5885205507392 200.275726439937
86.3316840701158 201.5756522997
87.0813004643988 202.887884576784
87.83742576378 204.212516870784
88.6001164849602 205.549643812893
89.369429635374 206.899361070239
90.1454227174503 208.26176535038
90.9281537329102 209.636954405953
91.7176811871033 211.025027039473
92.5140640933794 212.426083108294
93.3173619775006 213.840223529709
94.1276348820899 215.267550286203
94.9449433711192 216.708166430856
95.7693485344362 218.162176092879
96.6009119923309 219.629684483303
97.4396959001408 221.110797900802
98.2857629528971 222.605623737662
99.139176390011 224.114270485879
100 225.636847743402
};
\draw[red, thick, dashed] (axis cs:27.5772, 100) -- (axis cs:27.5772, 250);

\addplot[mark=*, red, mark size=2pt] coordinates {(27.5772, 114.821)};


\end{axis}

\end{tikzpicture}
    \vspace{-0.7cm}
    \caption{Plot of $\tau(\eta+W_\tau)$ versus $\tau$, with the optimal value $\tau^*=28.1392$ indicated in red.}
    \label{fig:tau}
\end{figure}
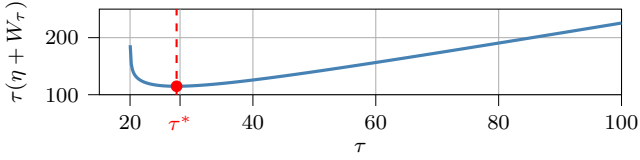

\begin{figure*}[!b] 
  \hrulefill 
  \begin{align}\label{eq:optVal}
     W_\tau &= \frac{1}{2\tau}x_\ini^\top\left( \Pi_0^{-1} - \frac{\Sigma_\ini}{\tau} \right)^{-1}x_\ini - \frac{1}{2}\logdet(\Sigma_\ini)-\frac{1}{2}\sum_{k=0}^{N-1} \log\left( \det \left(\Sigma_{k+1} \right) \det\left( P_k - \frac{C^\top C}{\underline{\sigma}^{2}} \right) \right) \\
     &-\frac{1}{2} \sum_{k=0}^{N-1} \logdet \left(I-\frac{K_k\left(\underline{\sigma}^2 I_p + C\left(P_k-\frac{C^\top C}{\underline{\sigma}^{2}}\right)^{-1}C^\top \right)K_k^\top \left(\Pi_{k+1}^{-1} - \frac{\Sigma_{k+1}}{\tau} \right)^{-1}}{\tau} \right) - \frac{1}{2}\logdet\left(\Sigma_{N}^{-1} - \frac{Q_{N}}{\tau} \right) \nonumber
  \end{align}
\end{figure*}

\begin{rem1}\label{rem:tradeoff}\upshape
    Unlike related works~\cite{HaleJone18,YazdJone22}, deriving an explicit tradeoff between privacy and control performance is nontrivial in our setting due to the minimax formulation. 
    In standard LQG, certainty equivalence ensures that noise statistics only affect the state estimator.
    In contrast, the privacy budget $\epsilon$ influences our proposed controller in a coupled manner. 
    Specifically, $\epsilon$ determines the lower bounds of $\underline{\sigma}^2$ and $\underline{b}$, which affects $\eta$. 
    This directly alters the optimal $\tau^*$ to \eqref{eq:equiv}, thereby impacting both the estimator and the feedback policy.
    A rigorous analysis on these effects is left for future work. 
\end{rem1}

\begin{figure}[t]
    \centering
        \begin{tikzpicture}
\definecolor{darkorange}{RGB}{255,127,14}
\definecolor{forestgreen}{RGB}{44,160,44}
\definecolor{mediumvioletred}{RGB}{197,33,156}
\definecolor{darkgray}{RGB}{176,176,176}

\pgfplotstableread{
x y1 y2 y3 y4 y5 y6
1.1920 49.77 42.72 90.47 99.27 200.91 231.52
1.2136 49.78 42.47 90.14 96.41 235.54 258.12
1.2353 50.25 42.94 90.93 97.27 209.59 247.87
1.2570 50.52 43.38 92.19 99.11 237.48 279.59
1.2787 50.36 43.18 91.49 97.37 252.82 284.83
1.3003 50.51 43.31 92.28 98.79 254.97 301.71
1.3220 50.51 43.06 90.97 97.21 214.19 245.22
1.3437 50.61 43.13 92.50 98.28 249.09 294.23
1.3653 50.43 43.08 90.12 96.40 247.44 290.01
1.3870 50.69 43.39 91.81 98.94 211.55 257.76
1.4087 50.85 43.55 92.40 98.48 254.37 304.70
1.4304 51.03 43.79 92.58 100.53 204.41 241.93
}\gaussianData

\pgfplotstableread{
x y1 y2 y3 y4 y5 y6
0.7213 49.58 42.31 89.83 97.15 255.19 303.49
0.7345 49.71 42.42 91.85 99.19 227.54 263.59
0.7476 49.42 42.50 91.19 99.33 210.54 236.15
0.7607 50.07 42.92 90.74 98.61 246.31 285.71
0.7738 49.62 42.41 91.85 97.15 246.55 290.43
0.7869 50.18 42.72 91.29 98.32 236.94 271.31
0.8000 50.54 43.18 92.51 99.41 201.84 241.93
0.8132 50.61 43.14 91.23 98.02 237.14 271.77
0.8263 50.86 43.29 92.55 98.55 259.06 310.85
0.8394 50.96 43.55 92.17 97.83 207.05 245.64
0.8525 51.26 44.03 92.77 98.31 249.12 299.12
0.8656 51.27 43.82 92.62 98.22 246.09 297.34
}\laplaceData

\newcommand{\myplots}[1]{
    \addplot [mediumvioletred, line width=1pt, mark=x, mark size=2.5] table[x=x, y=y1] {#1};
    \addplot [mediumvioletred, line width=1pt, dashed, mark=x, mark size=2.5, every mark/.append style={solid}] table[x=x, y=y2] {#1};
    \addplot [darkorange, line width=1pt, mark=*, mark size=1.5] table[x=x, y=y3] {#1};
    \addplot [darkorange, line width=1pt, dashed, mark=*, mark size=1.5,every mark/.append style={solid}] table[x=x, y=y4] {#1};
    \addplot [forestgreen, line width=1pt, mark=square*, mark size=1.5] table[x=x, y=y5] {#1};
    \addplot [forestgreen, line width=1pt, dashed, mark=square*, mark size=1.5,every mark/.append style={solid}] table[x=x, y=y6] {#1};
}

\begin{groupplot}[
    group style={
        group size=1 by 6,
        vertical sep=5pt, 
        group name=P
    },
    width=0.48\textwidth, height=0.15\textwidth,
    xmajorgrids, ymajorgrids, 
    x grid style={darkgray}, y grid style={darkgray},
    tick align=outside, tick pos=left,
    enlargelimits=false,
    xlabel style={align=center},
    label style={font=\footnotesize},      
    tick label style={font=\footnotesize}, 
    title style={font=\footnotesize},
    legend style={font=\footnotesize}
]


\nextgroupplot[
    ymin=200, ymax=310, ytick={200, 250, 300}, 
    xmin=1.18, xmax=1.45, 
    xticklabels={}, 
    axis x line* = top,
    legend entries={Mean (Proposed), Mean (LQG), 95th (Proposed), 95th (LQG), Worst (Proposed), Worst (LQG) }, 
legend columns=2,
transpose legend,
legend style={
    at={(0.5, 1.3)},       
    anchor=south,           
    font=\footnotesize, 
    /tikz/every even column/.append style={column sep=0.1cm},
    xshift=-0.5cm
}
]
\myplots{\gaussianData}

\nextgroupplot[
    ymin=90, ymax=105, ytick={90, 100}, 
    ylabel={$J(\mathcal{K})$},
    xmin=1.18, xmax=1.45, 
    xticklabels={}, 
    axis x line*=bottom,          
    x axis line style={draw=none}, 
    axis y line*=left,             
    execute at end axis={
        \draw[black] (rel axis cs:1,0) -- (rel axis cs:1,1);
    }
]
\myplots{\gaussianData}

\nextgroupplot[
    ymin=40, ymax=52, ytick={40, 50}, 
    xmin=1.18, xmax=1.45, 
    axis x line* = bottom, 
    xlabel={$\sigma^2$ \\ (a) Gaussian Mechanism},
    xlabel style={align=center}   
]
\myplots{\gaussianData}

\nextgroupplot[
ymin=200, ymax=320, 
ytick={200, 250, 300}, 
xmin=0.71, xmax=0.88, 
xticklabels={}, axis x line* = top,
yshift=-1.5cm
]
\myplots{\laplaceData}

\nextgroupplot[
ymin=88, ymax=102, 
ytick={90, 100}, ylabel={$J(\mathcal{K})$}, 
xmin=0.71, xmax=0.88,  
xticklabels={},
    axis x line*=bottom,          
    x axis line style={draw=none}, 
    axis y line*=left,             
    execute at end axis={
        \draw[black] (rel axis cs:1,0) -- (rel axis cs:1,1);
    }
]
\myplots{\laplaceData}

\nextgroupplot[ymin=40, ymax=52, ytick={40, 50}, 
xmin=0.71, xmax=0.88, 
axis x line* = bottom,
xlabel={$b$ \\ (b) Laplace Mechanism},
    xlabel style={align=center}  
]
\myplots{\laplaceData}

\end{groupplot}

\foreach \row in {1,2,4,5} {
    \foreach \side in {south west, south east} {
        \begin{scope}[shift={(P c1r\row.\side)}]
            \fill[white] (-4pt,-3pt) rectangle (4pt,3pt);
            \draw[thick] (-3pt,-3pt) -- (3pt,1pt);
            \draw[thick] (-3pt,-1pt) -- (3pt,3pt);
        \end{scope}
    }
}
\end{tikzpicture}
    \vspace{-0.6cm}
    \caption{Performance comparison of the proposed method and standard LQG over 10000 simulations. 
    The mean, $95$th percentile, and worst-case values of the cost $J(\mathcal{K})$ are plotted, while varying the noise parameters $\sigma^2$ and $b$ for (a) the Gaussian and (b) the Laplace mechanisms, respectively.}
    \label{fig:result}
\end{figure}
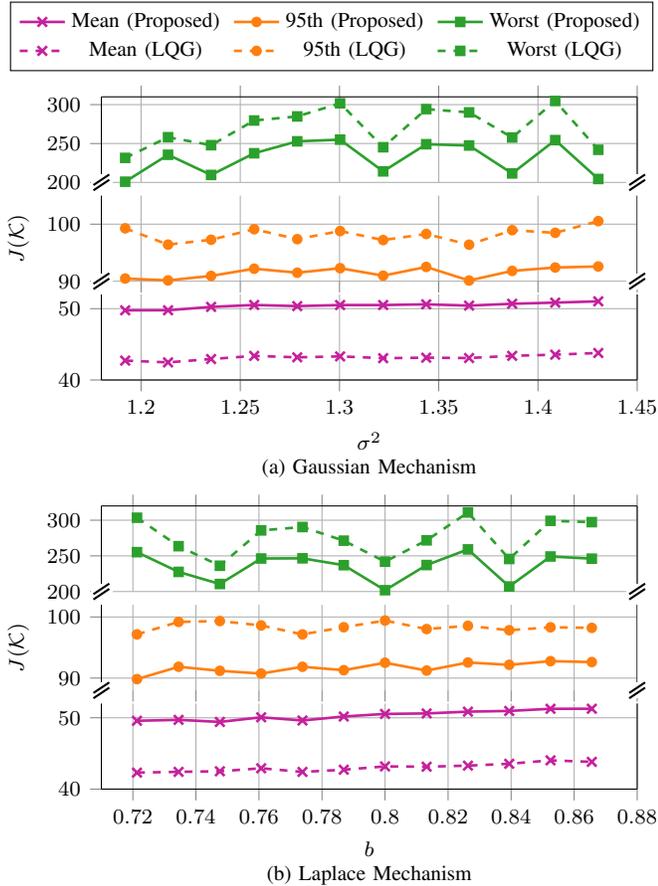

\section{Simulation Results}\label{sec:simul}

This section provides simulation results\footnote{Code fully available at https://github.com/yj-jang-98/DRO-DP} to demonstrate the effectiveness of the proposed method through a numerical example.
Consider the plant \eqref{eq:plant} given as
\begin{align*}
    A = 
    \begin{bmatrix}
        1.15 & 0.1 \\
        0 & 1.05
    \end{bmatrix}, 
    ~~~ 
    B = 
    \begin{bmatrix}
        1 \\ 0.5
    \end{bmatrix},
    ~~~
    C = 
    \begin{bmatrix}
        1 & 0,5
    \end{bmatrix},
\end{align*}
with $x_\ini = [1,-1]^\top$, $\Sigma_\ini=0.2 I_2$, and $\Sigma_w = 0.05I_2$.
We set the horizon length to $N=20$ and the weight matrices to $Q=Q_N=I_2$ and $R=0.3$.

We chose the DP and adjacency parameters as $(\epsilon,\delta) = (\ln(2),0.5)$ and $\gamma=0.5$, respectively.
The lower bounds for the noise parameters are set as $\underline{\sigma}^2 = 1.1920$ and $\underline{b}=0.7213$ to satisfy the conditions derived in Lemma~\ref{lem:mechGauss}. 
We empirically set the corresponding upper bounds to $\overline{\sigma}^2=1.2\underline{\sigma}^2$, and $\overline{b}=1.2\underline{b}$, and these parameter choices yield $\eta=1.8170$ according to Theorem~\ref{thm:ball}.
Fig.~\ref{fig:tau} depicts the value of $\tau(\eta+W_\tau)$ for different values of $\tau>0$. 
From this plot, we chose its optimal value as $\tau^*=28.1392$, with which the proposed controller is constructed based on Proposition~\ref{prop:controller}.

We compared the performance of the proposed method against a standard LQG controller designed under the assumption that $v_{0:N}\sim P_\nom$.
We selected the true noise parameter from a uniform grid over the admissible interval---
$[\underline{\sigma}^2,\overline{\sigma}^2]$ or $[\underline{b},\overline{b}]$ depending on the mechanism---containing $12$ points, and repeated the simulation $10000$ times for each chosen parameter. 
As shown in Fig.~\ref{fig:result}, the proposed method reduces both the $95$th percentile and the worst-case values of $J(\calK)$.
This can be thought of as a direct consequence of the risk-sensitive formulation derived in Theorem~\ref{thm:problEquiv}, which inherently assigns higher penalties to tail events.
These results suggest that the proposed method effectively achieves robustness against severe noise mismatches at the expense of a slight degradation in average-case performance.

\begin{figure}[t]
    \centering
        \input{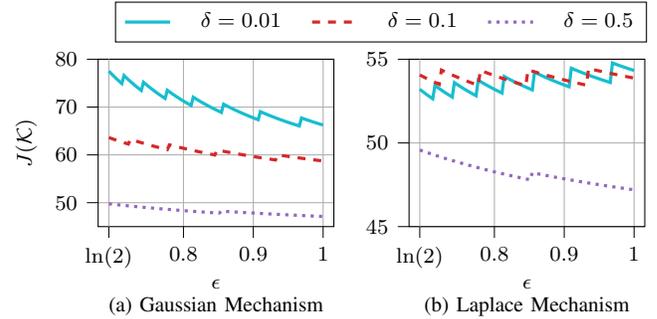}
    \vspace{-0.3cm}
    \caption{The cost $J(\calK)$ for different privacy parameters $(\epsilon,\delta)$, averaged over $10000$ simulations. 
    The ratio $\overline{\sigma}^2/\underline{\sigma}^2 = \overline{b}/\underline{b}=1.2$ is fixed for both the (a) Gaussian and (b) Laplace mechanisms.}
    \label{fig:privacy}
\end{figure}

Fig.~\ref{fig:privacy} illustrates the cost $J(\calK)$ of the proposed method under varying privacy parameters $(\epsilon,\delta)$, averaged over $10000$ simulations. 
For a fair comparison, $\underline{\sigma}^2$ and $\underline{b}$ are set as the lower bounds derived in Lemma~\ref{lem:mechGauss}, and the ratio $\overline{\sigma}^2/\underline{\sigma}^2 = \overline{b}/\underline{b}=1.2$ was fixed across all parameter sets. 
The results demonstrate a trend of performance degradation as privacy requirements increase, i.e., as $\epsilon$ decreases or $\delta$ decreases. 
However, it is not monotonic, possibly due to the coupled effects of privacy parameters discussed in Remark~\ref{rem:tradeoff}.

\section{Conclusion}\label{sec:conclusion}

In this paper, we have developed a distributionally robust optimal control approach for differentially private dynamical systems in which only an ambiguity set consisting of admissible noise distributions is known to the server. 
Accordingly, we formulated a minimax optimization problem to guarantee robust performance over the ambiguity set.
At the expense of suboptimality, we relaxed this ambiguity set into a convex KL divergence ball, so that the reformulated problem admits a tractable closed-form solution. 
Simulation results demonstrate that the proposed method achieves robust control performance against severe noise mismatches while ensuring DP.

\bibliographystyle{IEEEtran}
\bibliography{ref}

\appendix

\subsection{Proof of Lemma~\ref{lem:mechGauss}}\label{apdx:dp}
It follows from \cite[Theorem~A.1]{DworRoth14} that  the Gaussian mechanism satisfies $(\epsilon,\delta)$-DP if $\sigma^2 \ge 2\ln(1.25/\delta) \Delta_{2}^2/\epsilon^2,$ 
where $\Delta_{2}:=\sup_{(x_{0:N},x_{0:N}')\in\Adj}\|y_{0:N}-y'_{0:N}\|_2$.
Here $y_{0:N}$ and $y'_{0:N}$ correspond to the true output trajectories generated by $x_{0:N}$ and $x'_{0:N}$, respectively.
Similarly, the Laplace mechanism satisfies $\epsilon$-DP if $b \ge \Delta_{1}/\epsilon$ \cite[Theorem~3.6]{DworRoth14}, where $\Delta_{1}:=\sup_{(x_{0:N},x_{0:N}')\in\Adj}\|y_{0:N}-y'_{0:N}\|_1$.
Using the standard norm inequality $\|\cdot\|_2\le\|\cdot\|_1$ and \eqref{eq:adjDef}, we have
\begin{align*}
    \left\| y_{0:N}-y'_{0:N}\right\|_2 &\le  \left\| C \right\|_2 \left\| x_{0:N} - x'_{0:N} \right\|_2 \le \left\| C \right\|_2 \gamma, \\
    \left\| y_{0:N}-y'_{0:N}\right\|_1 &\le \left\|C \right\|_1 \gamma.
\end{align*}
Substituting these concludes the proof.


\end{document}